# Superconducting Vacuum-Gap Crossovers for High Performance Microwave Applications

Kevin L. Denis, Ari D. Brown, Meng-Ping Chang, Ron Hu,
Kongpop U-Yen, and Edward J. Wollack, *Senior Member IEEE*

*Abstract*— **The design and fabrication of low-loss wide-bandwidth superconducting vacuum-gap crossovers for high performance millimeter wave applications are described. In order to reduce ohmic and parasitic losses at millimeter wavelengths a vacuum gap is preferred relative to dielectric spacer. Here, vacuum-gap crossovers were realized by using a sacrificial polymer layer followed by niobium sputter deposition optimized for coating coverage over an underlying niobium signal layer. Both coplanar waveguide and microstrip crossover topologies have been explored in detail. The resulting fabrication process is compatible with a bulk micro-machining process for realizing waveguide coupled detectors, which includes sacrificial wax bonding, and wafer backside deep reactive ion etching for creation of leg isolated silicon membrane structures. Release of the vacuum gap structures along with the wax bonded wafer after DRIE is implemented in the same process step used to complete the detector fabrication.**

## I. INTRODUCTION

A low-loss crossover or "bridge" can be realized with a controlled-impedance structure spanning a transmission line. Such devices find utility in signal routing and suppression of parasitic excitations in single-mode planar circuitry [1,2]. Although a thin dielectric spacer can be used to provide signal isolation between the crossing lines, for low noise sensor and readout applications at sub-Kelvin temperatures a vacuum dielectric is preferable for the bridge both from microwave circuit and electromagnetic performance perspectives. Such orthogonal circuit crossings can have minimal coupling due to the reduced interaction area and improved modal symmetry [3,4]. Compensation of the junction reactance can be employed to further improve upon the achievable match, isolation, and signal bandwidth [5].

At radio and microwave frequencies aluminum wire bonds have been successfully employed to realize superconducting crossovers. There are a number of practical disadvantages of using this technique for the fabrication of complex circuits at millimeter wavelengths. Wire bonds require a large area relative to typical circuit features realized by photolithography and thus can lead to significant parasitic loading of the circuit. In practice, it is difficult to control the detailed geometry and thus the impedance of the wire bond. In addition, the placement and symmetry of such crossovers plays a key role in the suppression of spurious modes of signal propagation at high frequencies. If the bridge is absent or asymmetric with respect to the thru line, mode conversion can ultimately lead to relatively high losses, spurious resonances, and degradation of the circuit performance. As a result, for complex large-scale circuits this fabrication approach is not only challenging from a tolerance perspective, but presents yield and reliability concerns.

Although dielectric bridge structures can be envisioned which may be easier to fabricate – a vacuum gap is preferred in achieving a broadband response due to its lower capacitive parasitic loading and inter-signal line coupling. For low-noise applications this is also potentially desirable to avoid two-level system noise arising from charge fluctuations in amorphous dielectric materials. Via-less crossovers [6] can be realized as planar structures; however, this approach places limitations on the achievable bandwidth [7,8]. In response to these needs superconductive bridge crossovers with superconducting vias realized through standard planar sacrificial MEMS (Micro-Electro-Mechanical Systems) processes are explored in detail in this work. The resulting Nb transmission line structures are impedance matched and applicable for use as signal routing below the superconducting gap frequency (<0.7THz) in cryogenic millimeter wave sensors [9,10].

## II. FABRICATION PROCEDURE

### A. Crossover test structures

The yield and performance of simplified test structures were explored to verify the crossover fabrication process before incorporating in more complex settings. The process employed is similar to that employed in [11]. Direct current (DC) test structures for verifying superconducting continuity and mechanical properties, as well as microwave test structures incorporating single-mode CPW (co-planar waveguide) and microstrip line implementations were fabricated. The DC and CPW test structures do not require a separate isolated ground plane while the microstrip line implementation requires a dedicated ground plane for microwave evaluation. Single-crystal silicon is utilized as the

The authors gratefully acknowledge support from the NASA Astrophysics Research and Analysis (APRA) and the Goddard Space Flight Center Internal Research and Development (IRAD) programs.

Kevin L. Denis is with the NASA Goddard Space Flight Center, Greenbelt, MD 20771 USA (e-mail: kevin.l.denis@nasa.gov).
Ari D. Brown is with the NASA Goddard Space Flight Center, Greenbelt, MD 20771 USA Meng-Ping Chang is with SGT, Greenbelt, MD 20771 USA
Ron Hu is with STG, Greenbelt, MD 20771 USA, Kongpop U-Yen is with the NASA Goddard Space Flight Center, Greenbelt, MD 20771 USA, Edward J. Wollack is with the NASA Goddard Space Flight Center, Greenbelt, MD 20771 USA dielectric substrate for superconducting circuitry. The fabrication process utilizes a combination of Silicon-on-Insulator (SOI) wafer technology and polymer based wafer bonding in conjunction with a wafer-lapping step. This enables the SOI device layer to precisely define the microstrip dielectric substrate thickness. Further details are provided in [12,13,14]. Here the process described assumes the ground plane and microstrip dielectric, if needed, have already been fabricated. Note that the subsequent fabrication process steps are identical for both types (CPW or microstrip geometries) of crossover structures. The fabrication process steps are summarized in Fig 1.

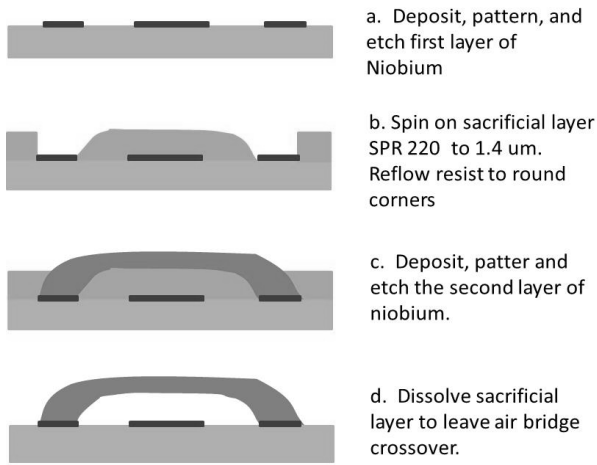

a. Deposit, pattern, and etch first layer of Niobium

b. Spin on sacrificial layer SPR 220 to 1.4 um. Reflow resist to round corners

c. Deposit, patter and etch the second layer of niobium.

d. Dissolve sacrificial layer to leave air bridge crossover.

Fig. 1. Fabrication process of niobium based crossovers.

The wafer is first coated with 300 nm of DC sputter deposited niobium (Nb1). The niobium is patterned by contact lithography and etched by reactive ion etching in a fluorine-based chemistry. The minimum feature size in this step is 2 μm. Next, the 1.4 μm sacrificial layer (Shipley SPR220-7 diluted with p-thinner) is spun on the wafer. The thickness varies by +/- 15 nm across the 80 mm active area of a 100 mm wafer. The choice of SPR resist was made due to the ease of reflowing the resist as well as it's compatibility with the subsequent sputter deposition and wax bonding necessary in the fabrication approach discussed below. The resist is then reflowed in an oven at 120 C for 40 min. The reflow step rounds the corners of the resist improving step coverage of the second niobium layer (Nb2).

The second layer of 0.5 μm thick niobium is deposited by a combination of DC sputter deposition and RF reverse bias etching of the wafer. This technique is known to improve step coverage of films due to the preferential etching rate of structures with sharp corners such as the photoresist sidewalls. The deposition is done in an AJA load locked system with a base pressure less than $1 \times 10^{-7}$ Torr. The process conditions are 250W DC power from a 3" niobium sputtering target and RF reverse bias etch at -10 W. The Ar pressure is held at 2.5 mTorr. The resulting film stress is slightly tensile 70 MPa. A tensile film stress is preferred to ensure that the released structures do not buckle after the sacrificial layer is removed.

The deposition rate is approximately 1 A/sec. After deposition Nb2 is patterned using the same SPR resist formulation to ensure compatibility. Nb2 is then reactive ion etched in CF4 and O2 chemistry with the photoresist serving as an etch stop. Finally, the photoresist sacrificial layer along with the Nb2 patterning layer are dissolved in acetone followed by a methanol rinse to release the vacuum-gap crossovers. Subsequent cleaning of the wafer by a low power oxygen plasma etching further cleans polymer residues. A final clean in Dupont PlasmaSolv EKC-265 followed by isopropyl alcohol rinse has been found to be highly effective in removing organometallic residue due to potential mixing of the Nb2 layer with the photoresist sacrificial layer. Immediately after the isopropyl alcohol rinse and without allowing the chip to dry a final methanol rinse is done followed by drying the chip in an oven at 120 C to complete the process.

## III. FABRICATION RESULTS

Fabrication design rules for the crossover were informed by making a set of test structure geometries. For the targeted applications, highly compact microstrip based crossovers designs are preferred in the microwave circuit layout. These geometries have lower complexity and lend themselves to higher yield due to their shorter span increased stiffness. The longest structure successfully yielded was the CPW test device, which spanned a length of 60 μm. See Fig. 2A. A typical failure mechanism for the CPW devices was buckling of the crossing line as shown in Fig. 2B. This failure mechanism can be attributed to stiction induced mechanical bonding of the niobium during the release process upon drying in the final methanol rinsing.

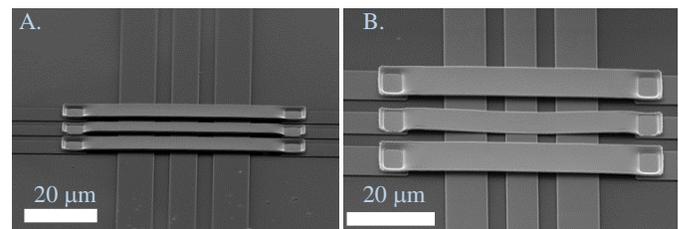

Fig. 2. CPW based vacuum-gap crossovers successfully released (A) and an example of center conductor bulking (B).

The yield was noted to improve for the longer structures through the use of a critical point drying process [15]. However, for the relatively shorter microstrip structures differences in yield were not apparent for either the methanol or critical point drying processes. A image of a chain of 264 crossovers in series is shown in Fig. 3 along with a DC measurement showing the superconducting transition with a critical temperature of 9.8 K and RRR = 2.66 at V = 30 μV. The via size for these structures is 10 μm x 10 μm. The vacuum-gap between the metal and the substrate is 1.3 μm and the length of the Nb2 layer between the substrate supports is 20 μm.

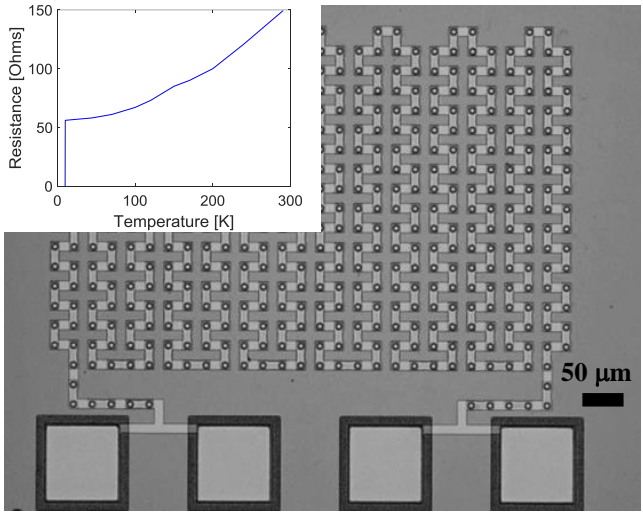

Fig. 3. Four wire crossover chain consisting of a series array of 264 niobium crossovers with inset showing superconducting transition of series array structure.

The simulated scattering-parameters as a function of frequency for the vacuum-gap microstrip are compared to via-less crossover design [6] in Fig 4. Ports 1 and 2 are connected to the bridge line and ports 3 and 4 are connected to the line orthogonal to the bridge, respectively. The characteristic impedance of the microstrip lines are 41 Ω. Low impedance line sections (31 Ω and have a ~8.5 degrees of electrical phase delay at 240 GHz) are placed on both sides of the bridge contacts to compensate for parasitic inductance introduced by the high characteristic impedance crossover structure. The bridge was made as narrow as practical to reduce parasitic coupling to the orthogonal line at the intersection. A bandwidth of more than 300 GHz and isolation of more than 27 dB was achieved using this approach. The structure can be placed in a sub-wavelength conductive enclosure to further reduce radiation losses as shown in Fig. 5.

## IV. DETECTOR PROCESS INTEGRATION

The properties a vacuum-gap crossover enable its use as a direct replacement for the via-less crossover in broadband polarimetric sensor systems. Vacuum-gap crossover devices were implemented in a 150 GHz planar orthomode transducer (OMT) architecture in a back-to-back test configuration. In this configuration, linearly polarized light is input aligned to one antenna polarization, propagates through the detector chip, and is measured through the orthogonal polarization with a vector network analyzer. An SEM photo of a completed chip is shown in Fig 6A. The vacuum-gap crossover is shown in Fig 6B. Two additional dummy crossovers are incorporated in the upper left of the image to maintain equal phase delay for each signal path in the OMT. The Nb1 and Nb2 layers are 4 μm wide at the crossover. The contact is 8.5 μm × 8.5 μm. The gap between the two layers is 1.2 μm, which is set by the thickness of the sacrificial layer. The total length of the Nb2 crossover is 10 μm.

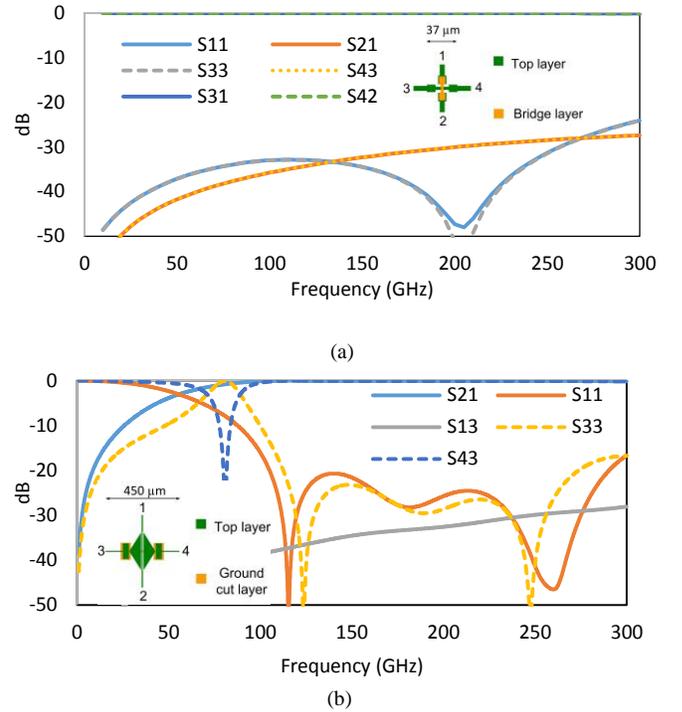

Fig. 4. Simulated S-parameter response of (a) vacuum bridge compared to (b) via-less crossover as a function of frequency. The via-less crossover signal crosses on the ground plane side from port 3 to 4.

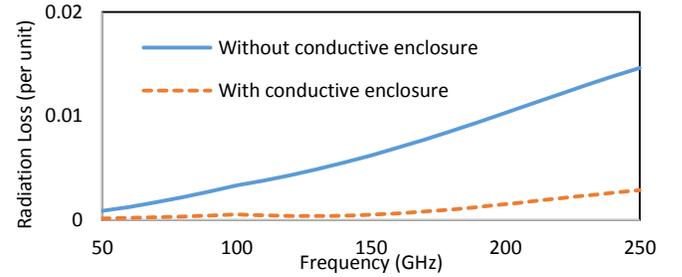

Fig. 5. Simulated radiation loss of vacuum-gap crossover with and without a conductive cavity enclosure.

The fabrication process employed is similar to that described in Fig 1. A significant difference is an additional wafer wax bonding step of the detector wafer to a temporary handle wafer and subsequent processing of the backside of the detector wafer. The wax bonding is incorporated after etching Nb2 and before removing the photoresist. This leaves the photoresist sacrificial layer under the crossover for mechanical support during wafer bonding. Crystal bond wax is used at 120 C with a Logitech wafer bonder. After wax bonding the backside of the detector wafer is patterned with 10 μm thick SPR220-7 photoresist and the silicon is etched by deep reactive ion etching (DRIE). The DRIE step is required to define the 5 μm thick silicon membrane supporting the antenna, and magic-tees. Future implementations will also include membrane isolated MoAu transition edge sensors. After the DRIE step, the resist and the polymer bonding material used in the SOI device layer transfer described above are removed in an oxygen plasma. Finally, the temporary handle wafer is removed by dissolving the wax in acetone.

The first fabrication run resulted in a mechanical crossover yield of approximately 96%. The microstrip crossovers did not have shorts between the Nb1 to Nb2 layers. A typical failure was mechanical damage possibly due to handling the structures during the final release step. Further work is warranted to improve the yield, but at this early stage in development this represents a promising result.

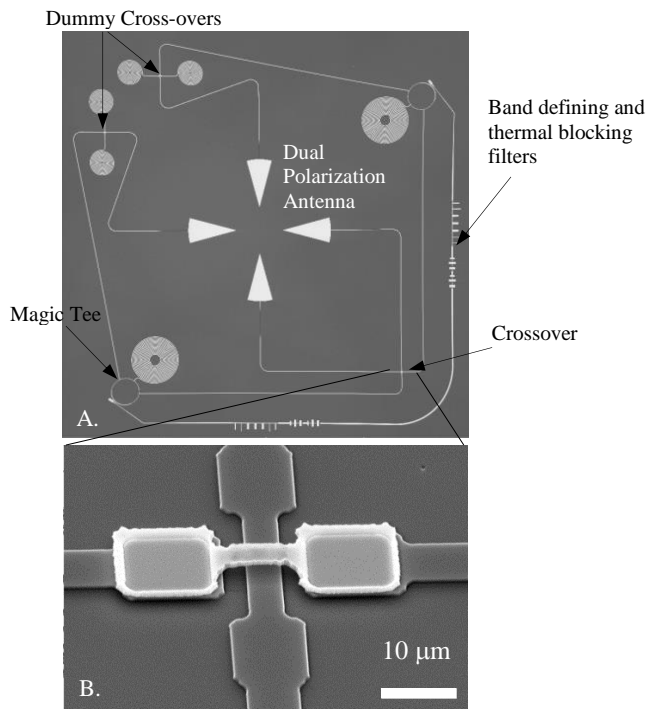

Fig. 6A. SEM of a 150 GHz dual polarization planar orthomode transducer incorporating vacuum-gap crossovers. The two dummy crossovers preserve signal phase presented to the magic-tee power 180° combiner. The spiral structures are normal metal broad-band load terminations. B. SEM photo of a crossover.

## V. Conclusion

A new fabrication process to integrate vacuum-gap crossovers in niobium metal with both superconducting microstrip and coplanar waveguide topologies is described. A surface micromachined fabrication process incorporating sacrificial polymer has been successfully integrated with a detector fabrication process incorporating a wax bonding step to a temporary handle wafer and a bulk micromachined deep reactive ion etching step followed by chip release. Due to the potential fragility of the vacuum-gap crossovers the process has been engineered to enable the final release and cleaning steps compatible with existing detector fabrication sequences. This process has been demonstrated in a millimeter wave orthomode transducer optimized for 150 GHz.